\documentclass[prl,twocolumn,showpacs]{revtex4}
\usepackage{graphicx}
\usepackage{amsmath}
\usepackage{bm}
\usepackage[T1]{fontenc}
\date{\today}
\begin{document}
\title{III-V and II-VI Mn-based Ferromagnetic Semiconductors} 
\
\author{Tomasz Dietl\footnote{Address in 2003: 
Institute of Experimental and 
Applied Physics, Regensburg University; supported by Alexander von Humboldt 
Foundation}}
\affiliation{Institute of Physics, Polish Academy of
Sciences, al.~Lotnik\'ow 32/46, 02-668 Warszawa, Poland}

\begin{abstract}
A review is given of advances in the field of carrier-controlled ferromagnetism in Mn-based diluted magnetic semiconductors and their nanostructures. Experimental results for III-V materials, where the Mn atoms introduce both spins and holes, are compared to the case of II-VI compounds, in which the Curie temperatures $T_{\mbox{\tiny{C}}}$ above 1 K have been observed for the uniformly and modulation-doped p-type structures but not in the case of n-type films. The experiments demonstrating the tunability of $T_{\mbox{\tiny{C}}}$ by light and electric field are presented. The tailoring of domain structures and magnetic anisotropy by strain engineering and confinement is discussed emphasizing the role of the spin-orbit coupling in the valence band. The question of designing modulated magnetic structures in low dimensional semiconductor systems is addressed. Recent progress in search for semiconductors with $T_{\mbox{\tiny{C}}}$ above room temperature is presented.
\end{abstract}
\maketitle

Over the recent years spin electronics (spintronics) has emerged as an interdisciplinary field of nanoscience, whose main goal is to acquire knowledge on spin-dependent phenomena, and to exploit them for new functionalities \cite{Spin01}. One of the relevant issues is to develop methods suitable for manipulations with the magnetization magnitude and direction as well as with the spin currents, which will ultimately lead to a control over the individual spins in solid state environment. Today's spintronic research involves virtually all material families, the most mature being studies on magnetic metal multilayers, in which spin-dependent scattering and tunnelling are being successfully applied in reading heads of high density hard-discs and in magnetic random access memories (MRAM). However, particularly interesting appear to be ferromagnetic semiconductors, which combine complementary functionalities of ferromagnetic and semiconductor material systems. For instance, it can be expected that powerful methods developed to control the carrier concentration and spin polarization in semiconductor quantum structures could serve to tailor the magnitude and orientation of magnetization produced by the spins localized on the magnetic ions. Furthermore, there is a growing amount of evidences that ferromagnetic semiconductors are the materials of choice for the development of functional spin injectors, aligners, filters, and detectors.  In addition of consisting the important ingredient of power-sawing spin transistors, spin injection can serve as a tool for fast modulation of light polarization in semiconductors lasers.

Already early studies of Cr spinels as well as of rock-salt Eu- \cite{Wach79} and Mn-based \cite{Stor97} chalcogenides led to the observation of a number of outstanding phenomena associated with the interplay between ferromagnetive cooperative phenomena and semiconducting properties. The discovery of the carrier-induced ferromagnetism in Mn-based zinc-blende III-V compounds \cite{Ohno92,Ohno96}, followed by the prediction \cite{Diet97} and observation of ferromagnetism in p-type II-VI materials \cite{Haur97,Ferr01} allows one to explore the physics of previously not available combinations of quantum structures and magnetism in semiconductors \cite{Ohno98}. These aspects of ferromagnetic semiconductors will be presented here together with a description of models aiming at explaining the nature of ferromagnetism in these materials. This survey has been prepared on the basis of recent review papers on ferromagnetic semiconductors \cite{Diet02,Mats02}, updated with some latest findings in this field. We limit ourselves only to Mn-based ferromagnetic semiconductors originating from zinc-blende family of diluted magnetic semiconductors (DMS). Other magnetic semiconductors like manganites, chalcogenides, and spinels, exhibiting different magnetic coupling mechanisms and band structures, are not discussed here.

\section{Mn impurity in II-VI and III-V Semiconductors}
\subsection{Substitutional Mn}
It is well established that substitutional Mn is divalent in II-VI compounds, and assumes the high spin d$^5$ configuration characterized by $S = 5/2$ and $g = 2.0$. Here, Mn ions neither introduce nor bind carriers, but give rise to the presence of the localized spins which are coupled to the effective mass electrons by a strong symmetry allowed p-d kinetic exchange and by a weaker s-d potential exchange \cite{Diet94}. In III-V compounds, in turn, the Mn atom, when substituting a trivalent metal, may assume either of two configurations: (i) d$^4$ or (ii) d$^5$ plus a weakly bound hole, d$^5$+h. It is now commonly accepted that the Mn impurities act as effective mass acceptors (d$^5$+h) in the case of antimonides and arsenides, so that they supply both localized spins and holes, a picture supported by MCD \cite{Szcz01} and EPR \cite{Fedo02} measurements. Just like in other doped semiconductors, if the average distance between the Mn acceptors becomes smaller than $2.5a_B$, where $a_B$ is the acceptor Bohr radius, the Anderson-Mott insulator-to-metal transition occurs. However, a strong p-d antiferromagnetic interaction between the Mn and hole spin enhances strongly the acceptor binding energy and reduces $a_B$. It has been postulated \cite{Diet02a} that owing to the large p-d interaction, the effect is particularly strong in nitrides, and may lead to the formation of a middle-gap small d$^5$+h polaron state, reminiscent of the Zhang-Rice singlet in high temperature superconductors. 

\subsection{Interstitial Mn}

Another important consequence of electrical activity of Mn in III-V compounds is the effect of self-compensation. In the case of (Ga,Mn)As, it accounts presumably for the upper limits of both hole concentration and substitutional Mn concentration \cite{Yu02}. According to RBS and PIXIE experiments \cite{Yu02}, an increase in the Mn concentration not only results in the formation of MnAs precipitates \cite{DeBo96} but also in the occupation by Mn of interstitial positions, Mn$_I$. Since the latter is a double donor in GaAs \cite{Mase01}, its formation is triggered by lowering of the system energy due to removal of the holes from the Fermi level. This scenario explains the reentrance of the insulator phase for large Mn concentrations \cite{Mats98} as well as a strong influence of (Ga,Mn)As properties upon annealing at temperatures much lower than those affecting other possible compensators, such as As antisites, As$_{Ga}$. Importantly, a symmetry analysis demonstrates that the hybridization between bands and d-states of Mn$_I$ is weak, which implies a substantial decrease of the sp-d exchange interaction once Mn assumes an interstitial position \cite{Blin03}. 

\section{Zener Model of Carrier-mediated Ferromagnetism}

For low carrier densities, II-VI DMS are paramagnetic but neighbor Mn-Mn pairs are antiferromagnetically coupled or even blocked owing to short-range superexchange interactions. However, this antiferromagnetic coupling can be overcompensated by ferromagnetic interactions mediated by band holes \cite{Diet97,Haur97,Ferr01}. In the presence of band carriers, the celebrated Ruderman-Kittel-Kasuya-Yosida (RKKY) mechanism of the spin-spin exchange interaction operates. In the context of III-V magnetic semiconductors, this mechanism was first discussed by Gummich and da Cunha Lima \cite{Gumm90}, and then applied to a variety of III-V and II-VI Mn-based layered structures \cite{Bose99}. 

\begin{figure}
\includegraphics[width=95mm]{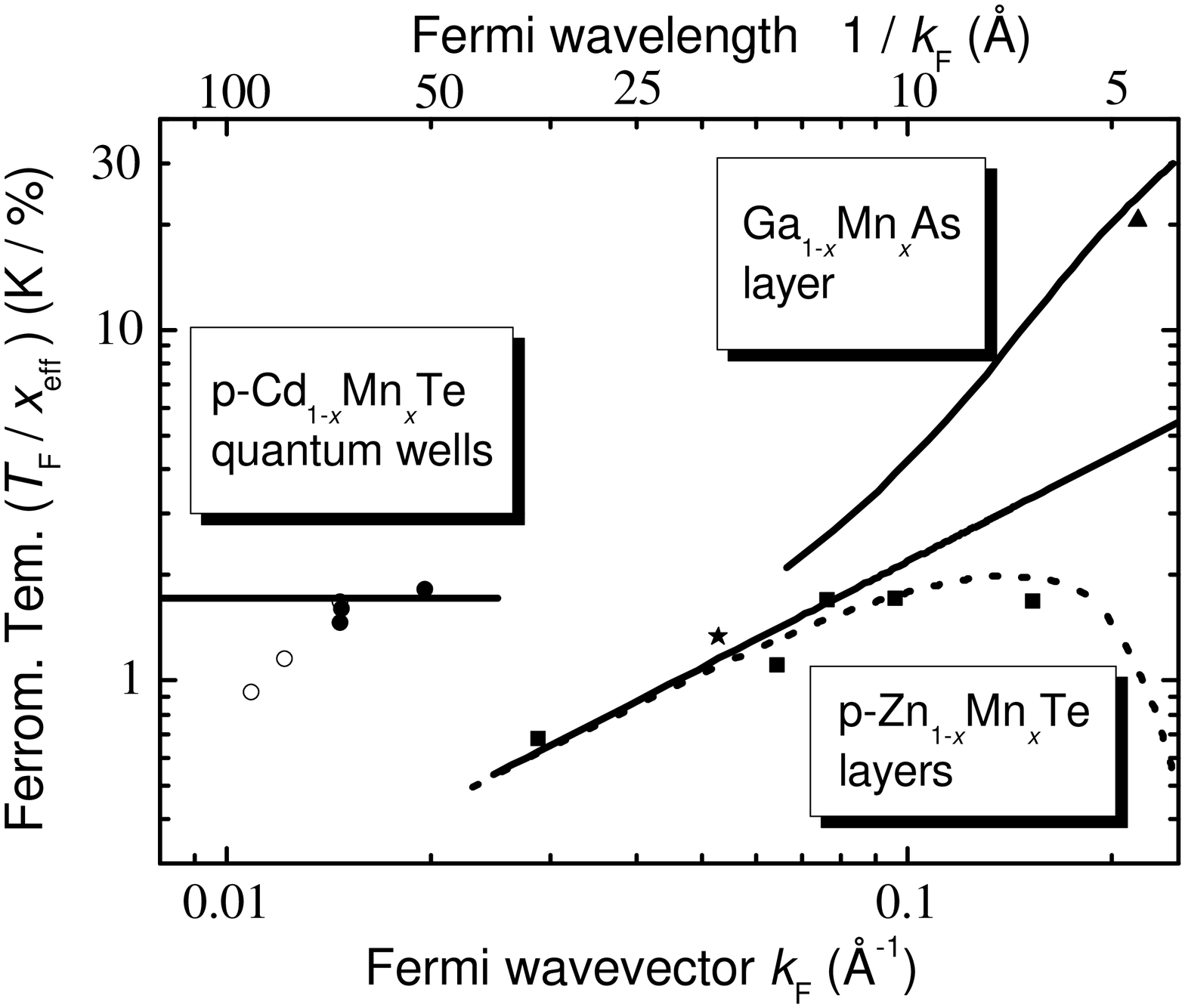}
\caption[]{Ferromagnetic Curie temperature normalized by the Mn concentration in epilayers of p-type (Zn,Mn)Te \cite{Ferr01,Andr01} and (Ga,Mn)As \cite{Mats98,Omiy00,Diet00,Diet01} as well as in modulation-doped p-type (Cd,Mn)Te \cite{Haur97,Bouk02}. Points and lines represent experimental and theoretical results, respectively}
\label{TC}
\end{figure} 

It has been shown \cite{Diet97} that on the level of the mean-field and continuous medium approximations, the RKKY approach is equivalent to the Zener model. In terms of the latter, the equilibrium magnetization, and thus $T_{\mbox{\tiny{C}}}$ is determined by minimizing the Ginzburg-Landau free energy functional $F[M(r)]$ of the system, where $M(r)$ is the local magnetization of the localized spins \cite{Diet00,Diet01}. This is a rather versatile approach, to which carrier correlation, confinement, $k\cdot p$, and spin-orbit couplings as well as weak disorder and antiferromagnetic interactions can be introduced in a controlled way, and within which the quantitative comparison of experimental and theoretical results is possible \cite{Ferr01,Diet01}.  As shown in Fig.~\ref{TC}, theoretical calculations \cite{Ferr01,Diet01}, carried out with no adjustable parameters, explain satisfactory the magnitude of $T_{\mbox{\tiny{C}}}$ in both (Zn,Mn)Te \cite{Ferr01} and (Ga,Mn)As \cite{Mats98,Omiy00}. A similar conclusion has been reached by analyzing $T_{\mbox{\tiny{C}}}$ in a series of annealed samples \cite{Edmo02}, in which $T_{\mbox{\tiny{C}}}$ reaches presently 160~K \cite{Edmo02a,Ku02,Ohno03}. In the model, the hole contribution to $F$ is computed by diagonalizing the $6\times 6$ Kohn-Luttinger $k\cdot p$ matrix containing the p-d exchange contribution, and by a subsequent computation of the partition function $Z$, $F_c = k_BT\ln Z$. The model is developed for p-type zinc-blende and wurzite semiconductors and allows for the presence of both biaxial strain and quantizing magnetic field. The enhancement of the tendency towards ferromagnetism by the carrier-carrier exchange interactions is described in the spirit of the Fermi liquid theory. Importantly, by evaluating $F_c(q)$, the magnetic stiffness can be determined, which together with magnetic anisotropy, yield the dispersion of spin waves \cite{Koen01} and the structure of magnetic domains \cite{Diet01a}. Owing to a relatively small magnitudes of the s-d exchange coupling and density of states, the carrier-induced ferromagnetism is expected \cite{Diet97} and observed only under rather restricted conditions in n-type DMS \cite{Andr01,Jaro02}.

\section{Spin Polarization}

\begin{figure}
\includegraphics[width=95mm]{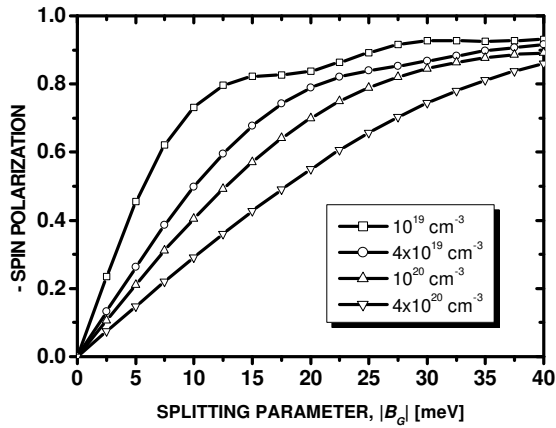}
\caption[]{Computed degree of spin polarization of the hole liquid as a function of the spin splitting parameter for various hole concentrations in  Ga$_{1-x}$Mn$_{x}$As ($B_G = -30$ meV corresponds to the saturation value of Mn spin magnetization for $x=0.05$). The polarization of the hole spins is oriented in the opposite direction to the polarization of the Mn spins (after Dietl {\it et al.}~\cite{Diet01})}
 \label{Spin_polarization}
\end{figure}

An important parameter that characterizes any magnetic material is the degree of spin polarization of band carriers. According to theoretical results \cite{Diet01} summarized in Fig.~\ref{Spin_polarization}, the expectation value of spin polarization reaches 80\% for typical values of Mn and hole concentrations in (Ga,Mn)As, a prediction that is being verified by Andreev reflection.

\section{Strain Effects}

\begin{figure}
\includegraphics[width=80mm]{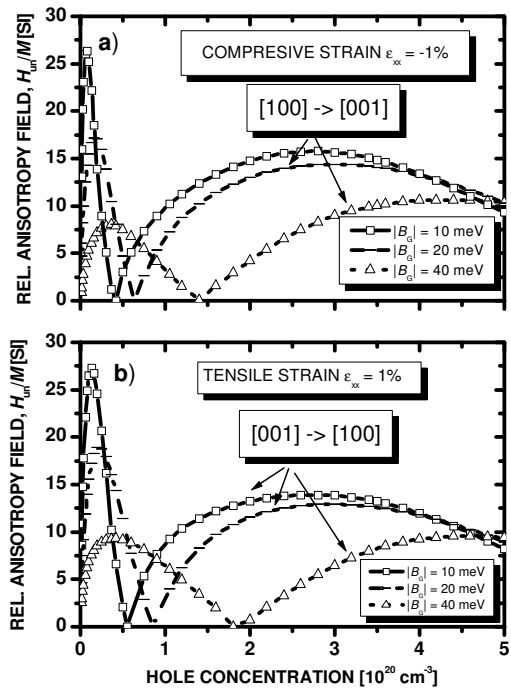}
\caption[]{Computed anisotropy field for compressive (a) and tensile (b) strains for various value of the hole spin splitting parameter $B_{\mbox{\tiny G}}$. The value of $B_{\mbox{\tiny G}} = 30$~meV corresponds to the saturation value of magnetization for Ga$_{0.95}$Mn$_{0.05}$As. The symbol [001] $\rightarrow$ [100] means that the easy axis is along [001], and the aligning external magnetic field is applied along [100] 
(after Dietl {\sl et al.}~\cite{Diet01})}
\label{PRB_strain}
\end{figure}

\begin{figure}
\includegraphics[width=90mm]{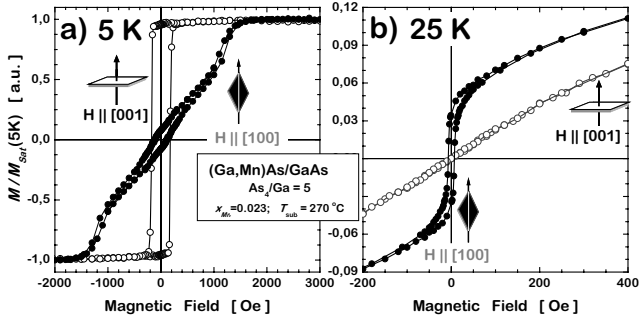}
\caption[]{Magnetization loops at 5~K (a) and 25~K (b) for parallel (full symbols) and perpendicular (open symbols) orientation of the (001) (Ga,Mn)As/GaAs epilayer with respect to the external magnetic field. The reversed character of the hysteresis loops indicates the flip of the easy axis direction between these two temperatures (after Sawicki {\sl et al.}~\cite{Sawi02})}
\label{Sawicki}
\end{figure}

It is well known that orbital momentum of the majority hole subbands depends on strain. Hence, magnetic anisotropy (easy axis direction) can be manipulated by adjusting the lattice parameter of the substrate, as the growth of DMS films in question is usually pseudomorphic. Theoretical results \cite{Diet01} displayed in Fig.~\ref{PRB_strain} show how magnetic anisotropy varies with the strain direction and the hole concentration \cite{Diet00,Diet01,Abol01}. In particular, the crystallographic orientation of the easy axis depends on whether the epitaxial strain is compressive or tensile, in agreement with the pioneering experimental studies for (In,Mn)As \cite{Mune93} and (Ga,Mn)As \cite{Ohno96a}. However,  magnetic anisotropy at given strain is predicted to vary with the degree of the occupation of particular hole subbands. This, in turn, is determined by the ratio of the valence band exchange splitting to the Fermi energy, and thus, by the magnitude of spontaneous magnetization, which depends on temperature. As shown in Fig.~\ref{Sawicki}, the predicted temperature-induced switching of the easy axis has recently been detected in samples with appropriately low hole densities \cite{Sawi02,Taka02}.

\section{Dimensional Effects}

It is straightforward to generalize the mean-field Zener model for the case of carriers confined to the $d$-dimensional space \cite{Diet97,Diet99}. The tendency towards the formation of spin-density waves in low-dimensional systems \cite{Diet99,Koss00} as well as possible spatial correlation in the distribution of the magnetic ions can also be taken into account. The mean-field value of the critical temperature $T_{{\bf q}}$, at which the system undergoes the transition to a spatially modulated state characterized by the wave vector ${\bf q}$, is given by the solution of the equation,
\begin{equation}
\beta^2A_F({\bf q},T_{{\bf q}})\rho_s({\bf q},T_{{\bf q}}) \int d
{\bf{\zeta}}\chi_o({\bf q},T_{{\bf q}},{\bf{\zeta}})
|\phi_o({\bf{\zeta}})|^4 = 4g^2\mu_B^2.
\end{equation}
Here ${\bf q}$ spans the $d$-dimensional space, $\phi_o({\bf{\zeta}})$ is the envelope function of the carriers confined by a $(3 - d)$-dimensional potential well $V({\bf{\zeta}})$; $g$ and $\chi_o$ denote the Land\'e factor and the ${\bf q}$-dependent magnetic susceptibility of the magnetic ions in the absence of the carriers, respectively. Within the mean-field approximation (MFA), such magnetization shape and direction will occur in the ordered phase, for which the corresponding $T_{{\bf q}}$ attains the highest value. A ferromagnetic order is expected in the three dimensional (3D) case, for which a maximum of $\rho_s({\bf q})$ occurs at $q = 0$.

According to the above model $T_{\mbox{\tiny{C}}}$ is proportional to the density of states for spin excitations, which is energy independent in the 2D systems. Hence, in the 2D case, $T_{\mbox{\tiny{C}}}$ is expected to do not vary with the carrier density, and to be enhanced over the 3D value at low carrier densities. Experimental results \cite{Haur97,Bouk02} presented in Fig.~\ref{TC} confirm these expectations, though a careful analysis indicates that disorder-induced band tailing lowers $T_{\mbox{\tiny{C}}}$ when the Fermi energy approaches the band edge \cite{Bouk02,Koss00}. In 1D systems, in turn, a formation of spin density waves with $q = 2k_F$ is expected, a prediction awaiting for an experimental confirmation. 

\section{Magnetization Manipulations by Electric Field and Light}

\begin{figure}
\includegraphics[width=90mm]{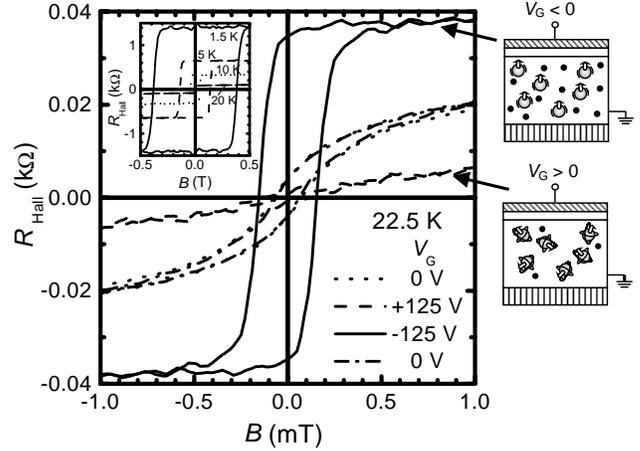}
\caption[]{Magnetization hysteresis evaluated by measurements of the anomalous Hall effect at various gate voltages that changes the hole concentration in a field-effect transistor structure with an (In,Mn)As channel (after Ohno {\it et al.}~\cite{Ohno00})}
\label{Ohno_Nature}
\end{figure} 

\begin{figure}
\includegraphics[width=95mm]{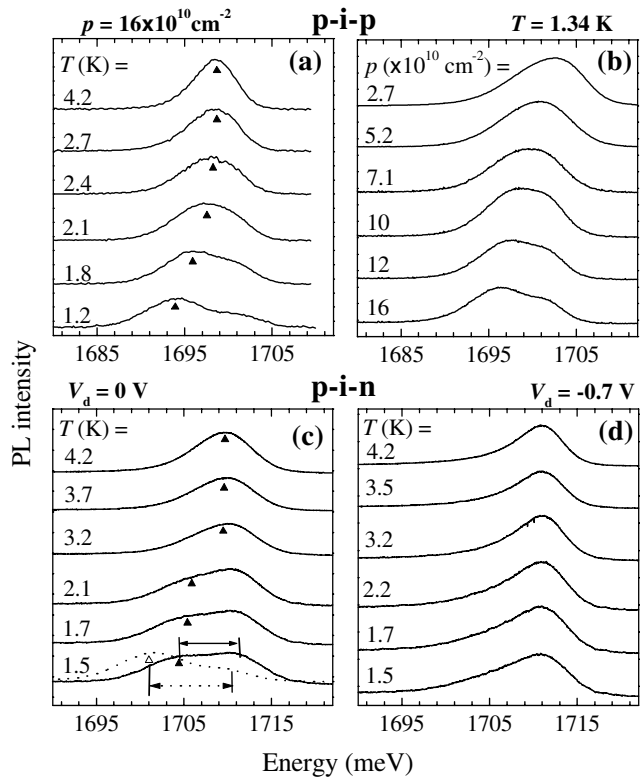}
\caption[]{Effect of temperature (a,c,d), illumination (b) and bias voltage $V_d$ (c,d) on photoluminescence line in quantum well of (Cd,Mn)Te placed in a center of p-i-n diode (c,d) and p-i-p structure (a,b). Line splitting and shift witness the appearance of a ferromagnetic ordering that can be altered isothermally, reversibly, and rapidly by light (b) and voltage (c,d), which change the hole concentration $p$ in the quantum well 
(after Boukari {\sl et al.}~\cite{Bouk02})}
\label{Boukari}
\end{figure}

Since magnetic properties are controlled by the band carriers, the powerful methods developed to change carrier concentration by electric field and light in semiconductor structures can be employed to alter the magnetic ordering. Such tuning capabilities of the materials in question were put into the evidence in (In,Mn)As/(Al,Ga)Sb \cite{Kosh97,Ohno00} and (Cd,Mn)Te/(Cd,Zn,Mg)Te \cite{Haur97,Bouk02} heterostructures, as shown in Figs.~\ref{Ohno_Nature} and \ref{Boukari}.  Importantly, the magnetization switching is isothermal and reversible. Though not investigated in detail, it is expected that underlying processes are rather fast. Since the background hole concentration is small in Mn-based II-VI quantum wells, the relative change of the Curie temperature is typically larger than in III-V compounds.

\section{Spin Injection}

\begin{figure}[htb]
\begin{minipage}[t]{90mm}
\includegraphics[width=80mm]{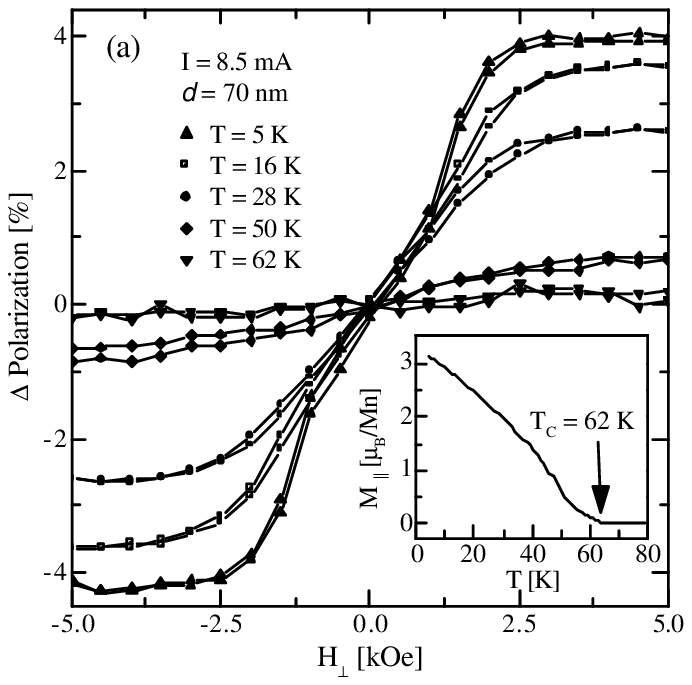}
\end{minipage}
\begin{minipage}[t]{60mm}
\includegraphics[width=40mm]{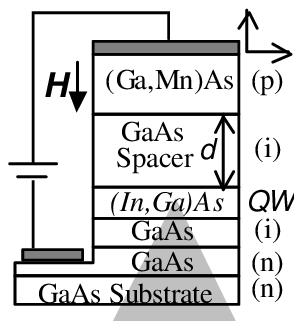}
\end{minipage}
\caption[]{Degree of circular polarization of light emitted by a (In,Ga)As quantum well located in a p-i-n diode biased in the forward direction and containing (Ga,Mn)As as the p-type electrode. An external magnetic field is applied along the hard axis, that is perpendicularly to the interface plane. The degree of circular polarization and the (Ga,Mn)As magnetization depend similarly on the magnetic field and temperature, which together with the lack of photoluminescence polarization for excitation by linearly polarized light, point to the existence of the hole spin injection from ferromagnetic (Ga,Mn)As to non-magnetic (In,Ga)As {\it via} a non-magnetic GaAs (after Young {\sl et al.}~\cite{Youn02})}
\label{Spin_injection}
\end{figure}

A number of groups is involved in the development of devices capable of injecting spins into a non-magnetic semiconductor.  Obviously, owing to a high degree of spin polarization and resistance matching ferromagnetic semiconductors constitute a natural material of choice here \cite{Ohno99}.  Typically, a pin light emitting diode structure is employed, in which the p-type spin injecting electrode is made of a ferromagnetic semiconductor. Experimental results obtained for the (Ga,Mn)As/GaAs/(In,Ga)As/n-GaAs diode are shown in Fig.~\ref{Spin_injection}. In this particular experiment \cite{Youn02}, the degree of circular polarization is examined for light emitted in the growth direction. In the corresponding Faraday configuration, simple selection rules are obeyed for radiative recombination between the electron and heavy hole ground state subbands. Since the easy axis is in plane, in agreement with the theoretical results of Fig.~\ref{PRB_strain}, a field of a few kOe is necessary to align the magnetization and thus to produce a sizable degree of light polarization.  

\section{Optical and Transport Properties}

In addition to thermodynamic properties discussed above, the Zener model of ferromagnetism in materials in question has been applied to describe optical properties of (Ga,Mn)As in the region of interband transitions between the exchange split valence band and the conduction band \cite{Szcz01,Diet01} as well in the regime, where intra-valence band excitations dominate \cite{Yang03}. Of course, both dispersion of absorption and magnetic circular dichroism are rather sensitive to disorder \cite{Szcz01,Yang03}, whose full description is difficult in these heavily doped and strongly disordered magnetic alloys. Furthermore, the so-far disregarded intra-d level transitions, enhanced presumably by the p-d hybridization, are expected to contribute in the spectral range in question. Nevertheless, the main aspects of available experimental results appear to be correctly understood.

Recently, various aspect of d.c. transport in (Ga,Mn)As and p-(Zn,Mn)Te have been examined. In general, a number of mechanisms by which the sp-d exchange interaction between localized and effective mass electrons can affect transport phenomena have been identified \cite{Diet94}. Generally speaking, these mechanisms are associated with spin-disorder scattering, spin-splitting, and the formation of bound magnetic polarons \cite{Diet94,Naga01}. In the particular case of (Ga,Mn)As, a resistance maximum at $T_{\mbox{\tiny{C}}}$ and the associated negative weak-field magnetoresistance have been described in terms of critical scattering \cite{Diet94,Omiy00} or by considering an interplay between the magnetic and electrostatic disorder \cite{Naga01,Yuld03}. Furthermore, anisotropic magnetoresistance (AMR), known already from early studies of p-(Hg,Mn)Te \cite{Wojt86}, has been theoretically examined as a function of strain and hole density in (Ga,Mn)As \cite{Jung02b}. Interestingly, it has been suggested \cite{Diet03} that the well-known high-field negative magnetoresistance of (Ga,Mn)As is actually an orbital weak localization effect, which is not destroyed by spin scattering owing to large spin splitting of the valence band.  Finally, arguments have been presented \cite{Jung02a} that owing to a relatively high resistance, the side-jump mmechanism of the anomalous Hall dominates, and its calculation with the appropriate Kohn-Luttinger amplitudes and by neglecting entirely the disorder gives a correct sign and amplitude of the Hall coefficient in both (Ga,Mn)As \cite{Jung02a} and p-(Zn,Mn)Te \cite{Diet03}.  

\section{Towards Functional Ferromagnetic Semiconductors}

\begin{figure}
\includegraphics[width=80mm]{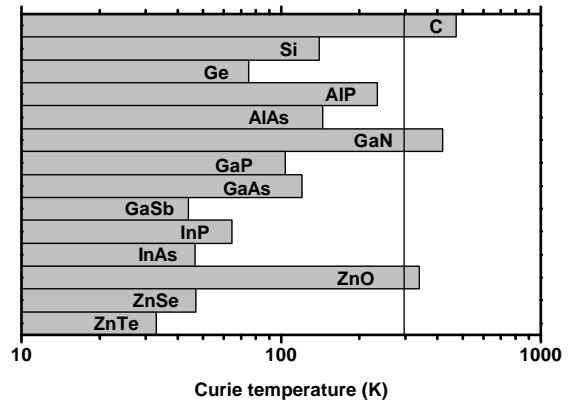}
\caption[]{Computed Curie temperature for various materials containing 5\% Mn per unit cell and $3.5\times 10^{20}$ holes per cm$^3$ (after Dietl {\sl et al.}~\cite{Diet00,Diet01})}
\label{T_C_Science}
\end{figure}

In view of the promising properties of ferromagnetic semiconductors, the development of a functional material with $T_{\mbox{\tiny{C}}}$ comfortably surpassing the room temperature, becomes an important challenge of today's  materials science.  A concentrated effort in this direction, stimulated by theoretical results \cite{Diet00,Diet01} recalled in Fig.~\ref{T_C_Science} and confirmed by others \cite{Jung02,Sato02}, suggests that there is no fundamental limits precluding the achievement of this goal. However, because of limited solubility of magnetic impurities in functional semiconductors, search for perspective compounds must be accompanied by a careful control and detection of possible ferromagnetic or ferrimagnetic precipitates and inclusions, typically with the sensitivity greater than that provided by standard x-ray diffraction. It is then useful to formulate some experimental criteria that should be fulfilled in order to call a given material a ferromagnetic semiconductor. First, magnetic characteristics should scale with the concentration of the magnetic constituent and also with the carrier density (which can be varied not only by doping but also by other means such as an electric filed or light). Furthermore, there should be a relation between temperature and field dependence of semiconductor and magnetic properties. In particular,  the anomalous Hall effect, spin-dependent resistance,  and magnetic circular dichroism together with the spin injection capability are the well known signature of a ferromagnetic semiconductor. 

\begin{figure}
\includegraphics[width=90mm]{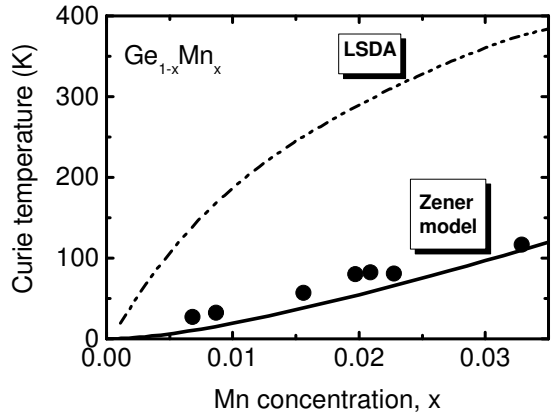}
\caption[]{Curie temperature in Ge$_{1-x}$Mn$_x$. Experimental results and LSDA theory are shown by points and dashed-dotted line respectively (after Joker {\sl et al.} \cite{Park02}). Solid line depicts the expectations of the Zener model \cite{Diet00,Diet01} assuming the hole concentration $p = 3.5\times 10^{20}$ [cm$^{-3}]x/0.025$}
\label{Science_letter}
\end{figure}

In the above context, particularly remarkable is a successful synthesis by the NRL group \cite{Park02} of a new ferromagnetic semiconductor Ge$_{1-x}$Mn$_x$. The epitaxial films of this material are shown to consist of (i) precipitates, whose size, composition, and magnetic properties depend on the growth temperature and (ii) a homogenous p-type matrix Ge$_{1-x}$Mn$_x$, whose $T_{\mbox{\tiny{C}}}$ increases approximately linearly with $x$, as shown in Fig.~\ref{Science_letter}. These findings are also compared with the outcome of theory developed within the local spin-density approximation (LSDA) \cite{Park02} and with the results determined from the Zener model of the hole-mediated ferromagnetism in tetrahedrally coordinated semiconductors, put forward a priori for Ge$_{1-x}$Mn$_x$ \cite{Diet00,Diet01}.  As seen, this comparison indicates that the LSDA overestimates considerably the magnitude of $T_{\mbox{\tiny{C}}}$, while the Zener model, despite the proximity of the metal-insulator transition, provides a reasonable evaluation of $T_{\mbox{\tiny{C}}}$ for the assumed degree of compensation.

There is a remarkable parallel effort aiming at synthesizing other promising ferromagnetic semiconductors. Limited ourselves to Mn-based III-V systems we can mention works demonstrating indications of high temperature ferromagnetism in (Ga,Mn)P \cite{Theo02} and (Ga,Mn)N \cite{Kuwa01,Sono02} though whether the ferromagnetism in these materials fulfilled the criteria specified above is under a vigorous debate \cite{Zaja01,Ando03}. 

\section{Conclusion}

With no doubt recent years have witness a remarkable progress in the development of new material systems, which show novel capabilities, such as the manipulation of ferromagnetism by the electric field. At the same time, carrier-controlled ferromagnetic semiconductors combine intricate properties of charge-transfer insulators and strongly correlated disordered metals with the physics of defect and band states in semiconductors. Accordingly, despite important advances in theory of these materials, quantitative understanding of these systems will be ahead for a long time.  

\section*{Acknowledgments}
The author would like to thank his co-workers, particularly F. Matsukura and H. Ohno in Sendai; J. Cibert in Grenoble; P. Kacman, P. Kossacki, and M. Sawicki in Warsaw, and A.H. MacDonald in Austin for many years of fruitful collaboration in studies of ferromagnetic semiconductors. Author's research in Germany in 2003 was supported by Alexander von Humboldt Foundation, while the work in Poland by State Committee for Scientific Research as well as by  FENIKS (G5RD-CT-2001- 00535) and AMORE (GRD1-1999-10502) EC projects.

\end{document}